\documentclass[aps,prl,preprint]{revtex4-1}
\usepackage{tikz}
\usepackage {graphicx}
\newcommand{\be}{\begin{equation}}
\newcommand{\ee}{\end{equation}}
\begin{document}
\title{A No-Boundary Method for Numerical Relativity}
\author{Lydia Bieri}
\email{lbieri@umich.edu}
\affiliation{Dept. of Mathematics, University of Michigan, Ann Arbor, MI 48109-1120, USA}
\author{David Garfinkle}
\email{garfinkl@oakland.edu}
\affiliation{Dept. of Physics, Oakland University, Rochester, MI 48309, USA}
\affiliation{Leinweber Center for Theoretical Physics, Randall Laboratory of Physics, University of Michigan, Ann Arbor, MI 48109-1120, USA}
\author{Shing-Tung Yau}
\email{yau@math.harvard.edu}
\affiliation{Dept. of Mathematics, Harvard University, Cambridge, MA 02138 USA}

\date{\today}

\begin{abstract}
We propose a method for numerical relativity in which the spatial grid is finite and no outer boundary condition is needed.  As a ``proof of concept'' we implement this method for the case of a self-gravitating, spherically symmetric scalar field.

\end{abstract}


\maketitle

\section{Introduction}

Asymptotically flat spacetimes are infinite in spatial extent, but computational grids are finite.  What then should a numerical relativity method do when simulating an asymptotically flat spacetime on a finite grid? In the asymptotic region, the metric behaves like a propagating wave on a flat background.  Thus the simplest thing to do is to place the outer boundary somewhat far out in the asymptotic region and to impose the sort of outgoing wave boundary condition that works for the wave equation.  However, such simple outer boundary conditions are not consistent with the Einstein field equation, and it is not clear whether the errors made by this inconsistency would be small.  The boundary, even if it is far out in the asymptotic region, encloses a finite region of space.  Therefore, from the mathematical point of view one should treat this situation by writing the Einstein field equation not as an initial value problem, but as an initial-boundary value problem, using only boundary conditions consistent with this sort of formulation.  Such an initial-boundary formulation was produced by Friedrich and Nagy in \cite{friedrichnagy}.  However, the formulation of \cite{friedrichnagy} is somewhat complicated and the variables it uses are not the sort usually used in numerical relativity (though see \cite{jorg1} for a numerical implementation).  Furthermore, it is not clear how within the allowed boundary conditions of \cite{friedrichnagy} to pick one that physically corresponds to outgoing gravitational waves.  

Another method is compactification at spatial infinity.\cite{meandcomer,frans}  Here one chooses spatial coordinates that make the outer boundary of the computational grid correspond to spatial infinity.  At that outer boundary, one simply sets the spatial metric to the Euclidean metric and the extrinsic curvature to zero.  This is consistent with the Einstein field equation. However eventually waves approach sufficiently close to the outer boundary that there are not enough grid points to resolve them and the simulation loses accuracy.  

To maintain resolution of the waves, one can instead compactify at null infinity.  One way to do this is through Cauchy-characteristic matching.\cite{winicour} Here each time slice consists of two pieces: a spacelike piece where variables are evolved using a standard Cauchy method, and a null piece where variables are evolved using a characteristic method.  The two sets of variables are matched at the place where the two parts of the surface join, and coordinates are chosen on the null piece so that the outer boundary of the grid is at null infinity.  Alternatively, the time slices can be ``hyperboloidal,'' that is spacelike slices that go out to null infinity.\cite{vince}  As yet another alternative, since the formalism of asymptotic flatness involves an unphysical spacetime conformally related to the physical spacetime, one can simply evolve the variables of the unphysical spacetime.\cite{friedrich,jorg2} Since in the unphysical spacetime, the extent of the physical spacetime is finite, it is natural to have the boundary of the computational grid correspond to the boundary of the physical spacetime within the unphysical spacetime.

In this paper, we propose a simple alternative to all these methods.  Our method involves only solving the Cauchy problem and does not require any specialized coordinates or any compactification methods.  The main idea of the method is illustrated in figure 1.  
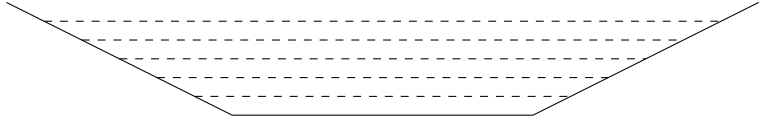
\begin{figure}
\begin{tikzpicture}
\draw (-3,1.5) -- (0,0) -- (4,0) -- (7,1.5);
\draw[dashed] (-0.5,0.25) -- (4.5,0.25);
\draw[dashed] (-1,0.5) -- (5.,0.5);
\draw[dashed] (-1.5,0.75) -- (5.5,0.75);
\draw[dashed] (-2,1) -- (6,1);
\draw[dashed] (-2.5,1.25) -- (6.5,1.25);
\end{tikzpicture} 
\caption{Slices for the No-Boundary method}
\label{fig:1}
\end{figure}
Here the solid lines represent the initial data and the dashed lines represent the constant time surfaces produced by the Cauchy evolution.  Note that the initial data consists of two pieces: a horizontal line representing the $t=0$ surface, and tilted lines that represent the rest of the initial data.  All the dashed lines are within the domain of dependence of the initial data.  In numerically evolving from one $t={\rm constant}$ surface to the next, most of the evolution is done using a standard Cauchy method, while the information for what to do at the boundary is provided by that part of the initial data on the tilted slices.  Since each dashed line is larger than the subsequent one, each numerical step involves adding points to the computational grid.  

As a proof of concept for this method, we will implement it for the case of a spherically symmetric, self-gravitating scalar field.  The equations of motion for this system are described in section 2, results of the simulations are presented in section 3, and a discussion is given in section 4.

\section{equations of motion}

We will choose the time slices to be maximal (trace of the extrinsic curvature vanishes) and the shift to be zero.  These conditions imply that the determinant of the spatial metric does not change with time, and we will choose coordinates on the initial slice so that the determinant is ${r^2} \sin \theta$.  With these conditions, the spacetime metric takes the form 
\be
d {s^2} = - {\alpha ^2} d {t^2} + {e^{-2A}} d {r^2} + {e^A} {r^2} (d {\theta ^2} + {\sin ^2}\theta d {\varphi ^2} ) \; \; \; .
\label{metric}
\ee
There is a scalar field $\Phi$ that satisfies the curved spacetime wave equation 
${\nabla ^a}{\nabla _a}\Phi=0$, and from the Einstein field equation we have 
\be
{R_{ab}} = {\nabla _a}\Phi {\nabla _b} \Phi
\label{Einstein}
\ee
where we are using units where $8\pi G=1$. 
We will put the wave equation in first order form by defining the quantities $P \equiv {n^a}{\nabla _a}\Phi$ and $\psi = {u^a}{\nabla _a}\Phi$ where $n^a$ is the unit normal vector to the $t= {\rm const.}$ hypersurfaces and $u^a$ is the unit radial vector in the hypersurface.  From eqn. (\ref{metric}) it then follows that
\begin{eqnarray}
P &=& {\alpha ^{-1}} {\frac {\partial \Phi}  {\partial t}} 
\label{P}
\\
\psi &=& {e^A} {\frac {\partial \Phi}  {\partial r}}
\label{psi}
\end{eqnarray}
From eqn. (\ref{P}) it follows that 
\be
{\frac {\partial \Phi}  {\partial t}} = \alpha P \; \; \; .
\label{dtPhi}
\ee
Taking the time derivative of eqn. (\ref{psi}) we obtain
\be
{\frac {\partial \psi}  {\partial t}} = {e^A} \left ( \alpha {\frac {\partial P} {\partial r}} + P {\frac {\partial \alpha}  {\partial r}} \right ) + \alpha {{K^r}_r} \psi \; \; \; .
\label{dtpsi}
\ee
Here ${K^r}_r$ is the eigenvalue of the extrinsic curvature in the radial direction, and we have used the fact that 
\be
{\partial _t} A = \alpha {{K^r}_r}
\label{dtA}
\ee
From the wave equation and equations (\ref{P}-\ref{psi}) we obtain
\be
{\frac {\partial P} {\partial t}} = {e^A} \psi {\frac {\partial \alpha}  {\partial r}} + {\frac \alpha  {r^2}} {\frac \partial {\partial r}} \left ( {r^2} {e^A} \psi \right ) \; \; \; .
\label{dtP}
\ee
Equations (\ref{dtPhi}), (\ref{dtpsi}) and (\ref{dtP}) constitute the equations of motion of our system.  However, in order to implement these equations we need to find the quantities $\alpha, \, A$ and ${K^r}_r$. 
The time components of the Einstein field equations yield a momentum constraint and a Hamiltonian constraint (Gauss-Codazzi equations). From the momentum constraint we obtain 
\be
{\frac {\partial {{K^r}_r}} {\partial r}} = - {{K^r}_r} \left ( {\frac 3  2} {\frac {\partial A}  {\partial r}} + {\frac 3  r} \right ) - {e^{-A}} P \psi \; \; \; ,
\label{momentum}
\ee
while the Hamiltonian constraint yields 
\be
{\frac {{\partial ^2}A}  {\partial {r^2}}} = {\frac 1  {r^2}} \left ( {e^{-3A}} - 1 \right ) - {\frac {\partial A}  {\partial r}} \left ( {\frac 5  r} + {\frac 7  4} {\frac {\partial A}  {\partial r}}\right ) - {e^{-2A}} \left [ {\frac 3 4} {{({{K^r}_r})}^2} + {\frac 1 2} ({P^2}+{\psi ^2}) \right ]  \; \; \; .
\label{Hamiltonian}
\ee
Finally, the maximal slicing condition yields 
\be
{\frac {{\partial ^2}\alpha } {\partial {r^2}}} + \left ( 2 {\frac {\partial A} {\partial r}} + {\frac 2 r} \right ) {\frac {\partial \alpha }  {\partial r}} = \alpha {e^{-2A}} \left [ {\frac 3 2} {{({{K^r}_r})}^2} + {P^2} \right ] \; \; \; .
\label{maximal}
\ee

The computer program works as follows: on a given time step, we know $P$ and $\psi$, and so our goal is to find $P$ and $\psi$ on the next time step.  We do this using eqns. (\ref{dtP}) and (\ref{dtpsi}), but in order to implement those equations, we need to find $A, \, {K^r}_r$ and $\alpha$.  We integrate eqns. 
(\ref{momentum}) and (\ref{Hamiltonian}) outward from $r=0$ using the boundary condition that $A$ and ${K^r}_r$ must be zero there.  Then we solve eqn. (\ref{maximal}) for $\alpha$ using the fact that 
$\partial \alpha /\partial r$ must vanish at $r=0$ and choosing the time coordinate so that $\alpha =1$ at the outer boundary.

Consistency of the equations requires that ${\partial _t} A - \alpha {{K^r}_r} =0 $ so we can use this as a check to see whether the code is working.  

However, we need one more piece of information to implement the evolution: the values of $P$ and $\psi$ at the outermost gridpoint of each time slice.  This information is given by initial data on the tilted part of the initial data slice, which we will refer to as the boundary surface.  Let ${\tilde n}^a$ be the unit normal to the boundary surface and let ${\tilde u}^a$ be the unit radial vector in the boundary surface.  In analogy with $P$ and $\psi$ define ${\tilde P} \equiv {{\tilde n}^a}{\nabla _a}\Phi$ and 
${\tilde \psi} \equiv {{\tilde u}^a}{\nabla _a}\Phi$.  Just as $P$ and $\psi$ can be freely specified on the horizontal part of the initial data surface, so $\tilde P$ and $\tilde \psi$ can be freely specified on the boundary surface, subject only to the condition that they match smoothly where the two parts of the initial data surface join.  
Since both $({n^a},{u^a})$ and $({{\tilde n}^a},{{\tilde u}^a})$ are orthonormal bases, there must be an angle $\beta$ such that 
\begin{eqnarray}
{{\tilde u}^a}={u^a}\cosh \beta + {n^a} \sinh \beta \; \; \; ,
\\
{{\tilde n}^a}={n^a}\cosh \beta + {u^a} \sinh \beta \; \; \; .
\end{eqnarray}
Inverting this relation, we find
\begin{eqnarray}
{u^a}={{\tilde u}^a}\cosh \beta - {{\tilde n}^a} \sinh \beta \; \; \; ,
\label{invertu}
\\
{n^a}={{\tilde n}^a}\cosh \beta - {{\tilde u}^a} \sinh \beta \; \; \; .
\label{invertn}
\end{eqnarray}
The boundary data that we need are the values of $P$ and $\psi$.  However, contracting eqns. (\ref{invertu}-\ref{invertn}) with 
${\nabla _a}\Phi$ we obtain
\begin{eqnarray}
\psi={\tilde \psi}\cosh \beta - {\tilde P} \sinh \beta \; \; \; ,
\label{boundpsi}
\\
P={\tilde P}\cosh \beta - {\tilde \psi} \sinh \beta \; \; \; .
\label{boundP}
\end{eqnarray}
Thus we can determine the boundary values of $P$ and $\psi$ from the boundary data 
$({\tilde P},{\tilde \psi})$ provided that we know the angle $\beta$.  We will choose the boundary surface to be generated by outgoing radial spacelike geodesics.  This condition along with the maximal slicing condition yields the following evolution equation for $\beta$.  
\be
{\frac {d\beta} {dt}}  =  {\frac {\alpha {{K^r}_r}} {\tanh \beta}} \; - \; {e^A} {\frac {\partial \alpha} {\partial r}}  \; \; \; .
\label{dtbeta}
\ee 
Thus at each time step, we evolve $\beta$ using eqn. (\ref{dtbeta}) and we use eqns. (\ref{boundP}) and (\ref{boundpsi}) to find $P$ and $\psi$ at the last gridpoint.
  
\section{results}

We choose the flat part of the initial data surface to be $t=0, \, 0 \le r \le {r_0}$ and the tilted part of the initial data surface to be $ t= (r-{r_0})/2, \, {r_0} \le r \le {r_{\rm max}}$ where 
$r_0$ and $r_{\rm max}$ are constants.  A simple way to get initial data that is smooth where the two parts of the surface join is simply to take a smooth function on spacetime and pull it back to the initial data surface.  We choose this spacetime function to be a solution of the flat spacetime wave equation. This is also a physically reasonable choice since far from the center a solution of the curved spacetime wave equation is well approximated by a solution of the flat spacetime wave equation.  In flat spacetime a smooth solution of the spherically symmetric wave equation takes the form 
\be
\Phi(t,r) = {\frac 1 r} ( f(t+r) - f(t-r))
\ee
where $f(s)$ is any smooth function.  We choose
\be
f(s) = {\frac a {1+ k {s^2}}}
\ee
Where $a$ and $k$ are constants.  This function represents a scalar field with amplitude $a$, with its energy concentrated around the center in a region of radius approximately $1/\sqrt k$.  Note that the initial value of $\Phi$ is zero, but its initial time derivative is nonzero.  We can check to see when a black hole forms by looking for a marginally outer trapped surface (MOTS).  The condition for a MOTS is
\be
1 + {\frac r 2} \left ( {\frac {\partial A} {\partial r}} + {e^{-A}} {{K^r}_r} \right ) = 0.
\ee
 
\begin{figure}
\centering
\includegraphics[width=4.5in]{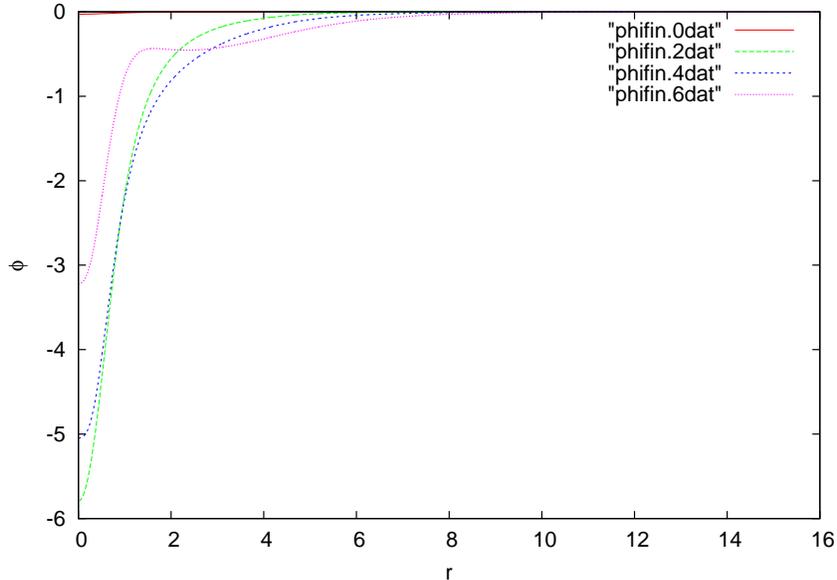}
\caption{$\Phi(r)$ at times 0, 2, 4 and 6.  Here the amplitude is 1.4 and no black hole forms}
\label{Phifig1}
\end{figure}

\begin{figure}
\centering
\includegraphics[width=4.5in]{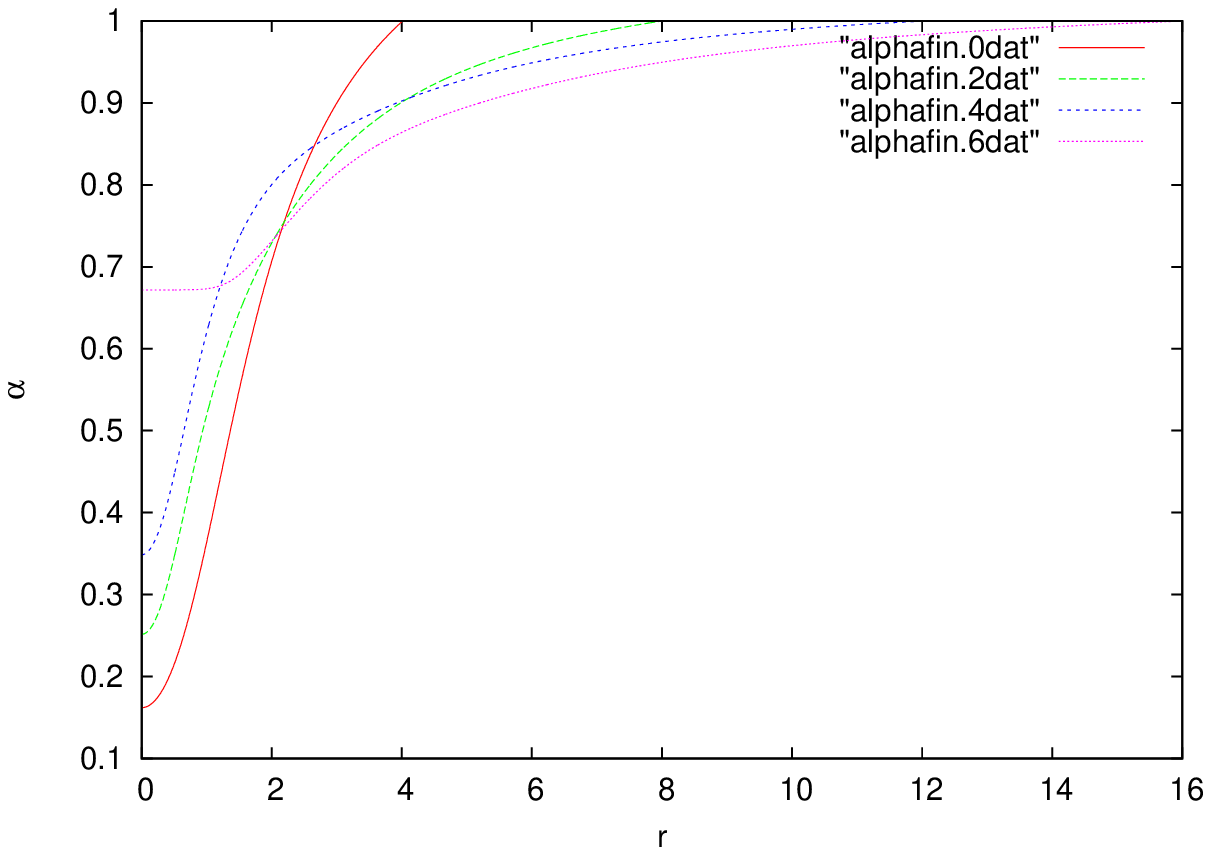}
\caption{$\alpha(r)$ at times 0, 2, 4 and 6.  Here the amplitude is 1.4 and no black hole forms}
\label{alphafig1}
\end{figure}

In figures (\ref{Phifig1}) and (\ref{alphafig1}) we show the results of a simulation where a black hole does not form.  In this case the scalar field eventually disperses and the lapse evolves towards its flat space value of 1.

\begin{figure}
\centering
\includegraphics[width=4.5in]{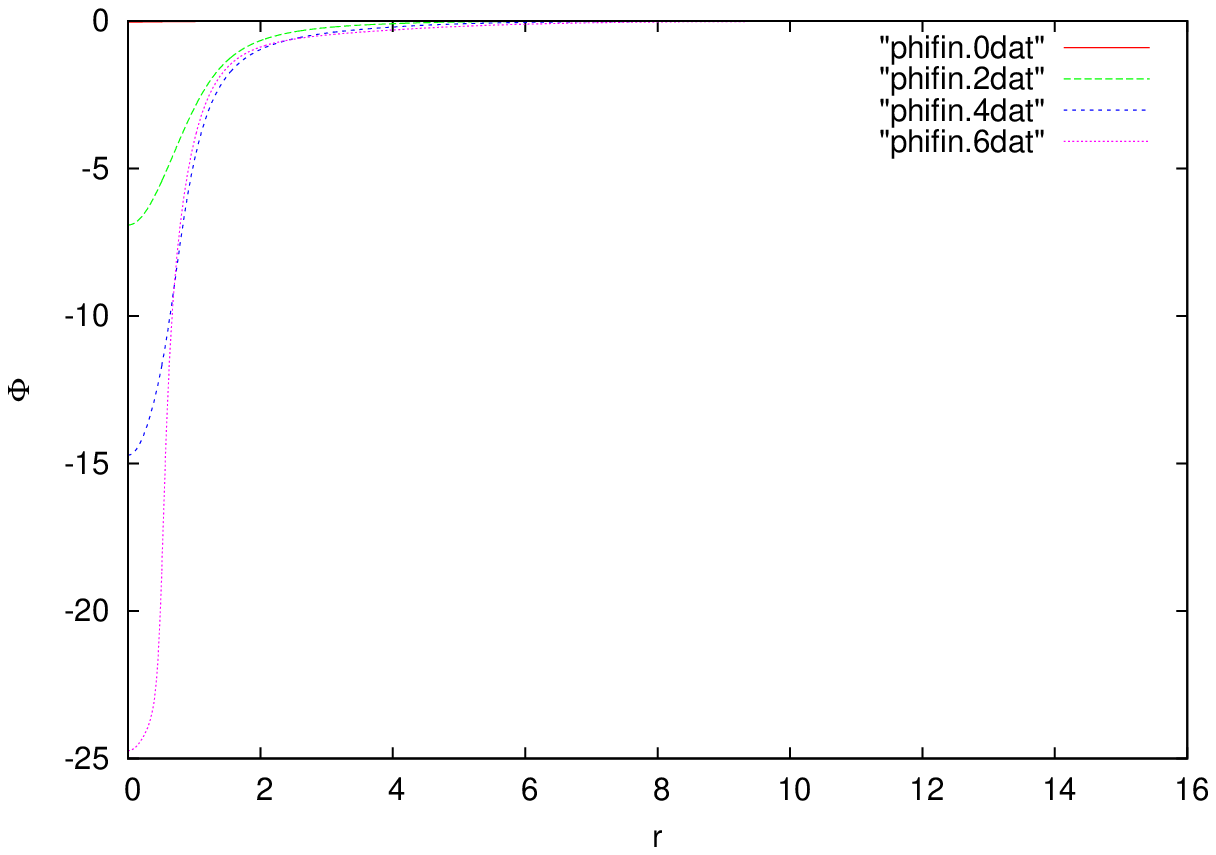}
\caption{$\Phi(r)$ at times 0, 2, 4 and 6.  Here the amplitude is 1.7 and a black hole forms}
\label{Phifig2}
\end{figure}

\begin{figure}
\centering
\includegraphics[width=4.5in]{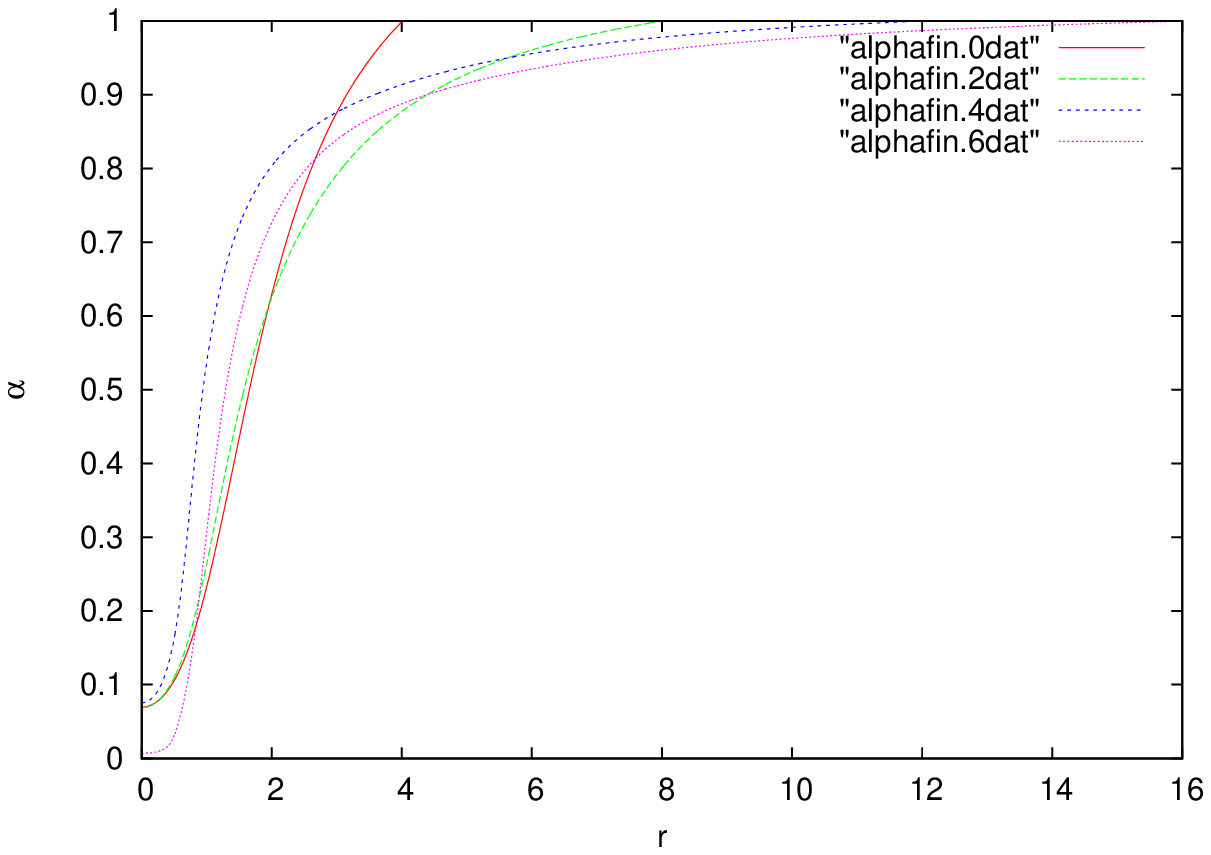}
\caption{$\alpha(r)$ at times 0, 2, 4 and 6.  Here the amplitude is 1.7 and a black hole forms}
\label{alphafig2}
\end{figure}

In figures (\ref{Phifig2}) and (\ref{alphafig2}) we show the results of a simulation where a black hole forms.  Here a portion of the scalar field remains trapped in the central region.  Also in this region there is the standard ``collapse of the lapse'' in which $\alpha$ takes on values close to zero.

\begin{figure}
\centering
\includegraphics[width=4.5in]{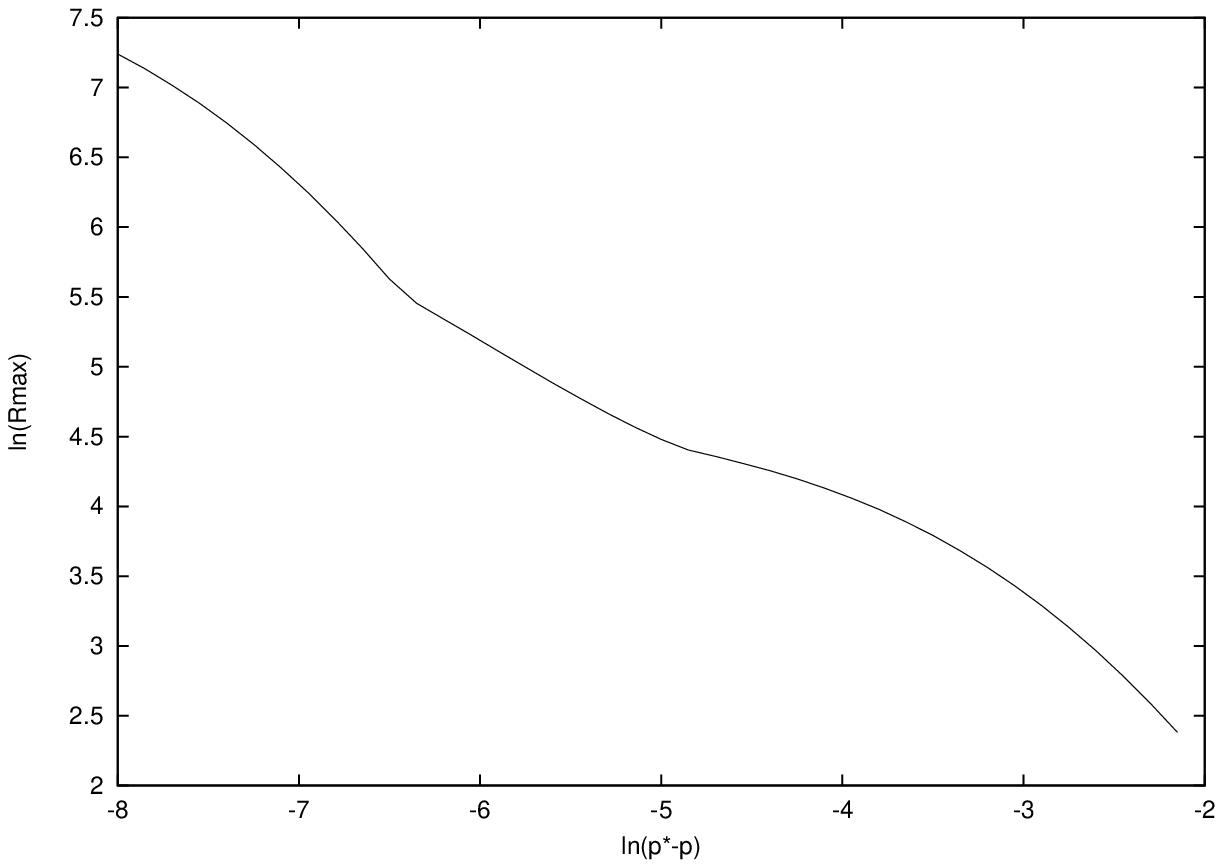}
\caption{curvature scaling for critical collapse}
\label{critfig}
\end{figure}

With our method, the code can only run for as long as there is data on the tilted part of the initial data surface.  But that is long enough, provided that all the interesting physics happens by that time.  To illustrate this point, we performed numerical simulations of critical gravitational collapse.\cite{choptuik} In critical collapse, one examines a family of initial data depending on a parameter $p$ where the threshold of black hole formation occurs at $p=p*$.  For $p$ slightly greater than $p*$, there is a scaling relation for black hole mass
\be
M \propto {{(p-p*)}^\gamma}
\label{massscale}
\ee
While for $p$ slightly less than $p*$ there is a scaling relation for the maximum value of the spacetime curvature\cite{meandcomer2}
\be
{R_{\rm max}} \propto {{({p*}-p)}^{-2\gamma}}
\label{curvscale}
\ee
where the constant $\gamma$ in eqn. (\ref{curvscale}) is the same as in eqn. (\ref{massscale}).  More precisely: eqn. (\ref{massscale}) implies that a graph of $\ln M$ vs. $\ln(p-p*)$ is a straight line with slope $\gamma$, while the actual result is a line with average slope $\gamma$ but with a small periodic wiggle.\cite{gundlach}  Correspondingly, in the subcritical case, a graph of $\ln (R_{\rm max})$ vs. $\ln(p*-p)$ is a line with average slope $-2\gamma$ but with a periodic wiggle. In our case, we use the amplitude $a$ as our parameter $p$.  Using a binary search, we find the critical value $p*$.  We then run a sequence of simulations slightly below the threshold of black hole formation.  The results are plotted in figure (\ref{critfig}).  Note that the maximum curvature has the appropriate scaling.  

Our critical collapse simulations were run with a maximum radius of 30 and with a fixed mesh size corresponding to a total of 50,000 grid points on the initial data surface. We thus have a fairly low resolution treatment of critical collapse, as compared to the higher resolution that can be obtained with mesh refinement.  Nonetheless, it is of interest to treat critical collapse with the maximal slicing condition that we use, because such slicing, in contrast to the slicing condition used in\cite{choptuik}, can follow the evolution even after the formation of a black hole.  This issue will be addressed elsewhere\cite{bruna} using mesh refinement for high resolution.   

\section{Discussion}

We now consider the possibility of generalizing our method to the case of the Einstein field equations with no symmetry.  The main features of our method should apply to any system of hyperbolic equations, and should therefore be suitable for Einstein's equations in the generalized harmonic formulation\cite{friedrichgh,megh,frans}
or the BSSN formulation\cite{shibata,thomas}.  In the spherically symmetric case, the initial data for the scalar field could be freely specified, while in the general case, the initial data for the Einstein field equations would have to satisfy constraint equations. However, this is no different from the usual Cauchy problem for the Einstein equation: the only difference is that the constraint equations would have to be solved on both the flat and tilted parts of the initial data surface, with care taken to impose a condition of smoothness at the place where the two parts of the initial data surface join.

Finally, we want to consider the possible expense, in terms of computer memory and time, of our method.  Since the method adds extra spatial points at each time step, it is possible that the method could become unweildy when run long enough to extract the relevant physics.  This could certainly be the case when using a Cartesian coordinate system, since in that case extra points need to be added in all three directions.  However, for a method using spherical coordinates\cite{thomas2} or where the outermost coordinate patch uses spherical coordinates\cite{sxs} one only needs to add radial points: no additional angular resolution is needed.  Furthermore, using our method it may well be that one could get away with making the initial outer boundary radius $r_0$ smaller than that of the fixed outer boundary used in the usual Cauchy codes.  Thus our method could be \emph{less} expensive at early times and only become more expensive towards the later parts of the simulation.  We therefore expect that our no-boundary method will be a useful addition to the tools used in numerical relativity.

\section*{Acknowledgments}

David Garfinkle thanks the Harvard Black Hole Initiative for hospitality, and acknowleges support from NSF Grants PHY-15105565 and PHY-1806219. Lydia Bieri acknowleges support from NSF grants DMS-1253149 and DMS-1811819 and Simons Fellowship in Mathematics 555809.  Shing-Tung Yau acknowledges support from NSF grant DMS-1607871.


\begin{thebibliography}{}




\bibitem{friedrichnagy}
H. Friedrich and G. Nagy, Commun. Math. Phys. {\bf 201}, 619 (1999)

\bibitem{jorg1}
J. Frauendiener and C. Stevens, Phys. Rev. D {\bf 89}, 104026 (2014)

\bibitem{meandcomer} 
D. Garfinkle and G. C. Duncan, Phys. Rev. D {\bf 63}, 044011 (2001)

\bibitem{frans}
F. Pretorius, Class. Quantum Grav. {\bf 23}, S529 (2006)

\bibitem{winicour}
J. Winicour, Living Rev. Relativ. (2012) 15:2.https://doi.org/10.12942/Irr-2012-2

\bibitem{winicour2}
N. T. Bishop, R. Gomez, L. Lehner, M. Maharaj, and J. Winicour, Phys. Rev. D {\bf 56}, 6298 (1997)

\bibitem{friedrich}
H. Friedrich, Commun. Math. Phys. {\bf 91}, 445 (1983)

\bibitem{jorg2}
J. Frauendiener, Phys. Rev. D {\bf 58}, 064002 (1998)

\bibitem{vince}
V. Moncrief and O. Rinne, Class. Quantum Grav. {\bf 26}, 125010 (2009)

\bibitem{choptuik}
M. W. Choptuik, Phys. Rev. Lett. {\bf 70}, 9 (1983)  

\bibitem{meandcomer2}
D. Garfinkle and G. C. Duncan, Phys. Rev. D {\bf 58}, 064024 (1998)

\bibitem{gundlach}
C. Gundlach, Phys. Rev. D {\bf 55}, 695 (1997)

\bibitem{bruna}
L. Bieri, D. Garfinkle, and B. Sanchez, in preparation

\bibitem{friedrichgh}
H. Friedrich, Commun. Math. Phys. {\bf 100}, 525 (1985)

\bibitem{megh}
D. Garfinkle, Phys. Rev. D {\bf 65}, 044029 (2002)

\bibitem{shibata}
M. Shibata and T. Nakamura, Phys. Rev. D {\bf 52}, 5428 (1995)

\bibitem{thomas}
T. W.  Baumgarte and S. L. Shapiro, Phys. Rev. D {\bf 59}, 024007 (1999)

\bibitem{thomas2}
T. W. Baumgarte, P. J. Montero, I. Cordero-Carrion, and E. Muller, Phys. Rev. D {\bf 87}, 044026 (2013)

\bibitem{sxs}
A. H. Mroue {\it et at} Phys. Rev. Lett. {\bf 111}, 241104 (2013) 


\end{thebibliography}
\end{document}